\documentclass[conference]{IEEEtran}

\usepackage{algorithm}
\usepackage{algorithmic}
\usepackage{mathtools}
\usepackage{cite}
\usepackage{hyperref}
\usepackage{amsmath}
\usepackage{algorithmic}
\usepackage{stfloats}
\usepackage{url}
\usepackage{times, epsfig}
\usepackage{fancyhdr}
\usepackage{amsfonts}
\usepackage{amssymb}
\usepackage{array}
\usepackage{graphicx}
\usepackage{url}
\usepackage{subfigure}
\usepackage{bm}
\usepackage{breqn}
\usepackage{xcolor}
\usepackage{soul}
\usepackage{amsthm}
\usepackage{todonotes}
\usepackage [utf8]{inputenc}
\usepackage{pdfcomment}

\begin{document}

\begin{titlepage}
\textbf{Title: TOD: Transprecise Object Detection to
Maximise Real-Time Accuracy on the Edge.}
\\Authors: JunKyu Lee, Blesson Varghese, Roger Woods, Hans Vandierendonck.
\\
\\Conference: Accepted at IEEE 5th International Conference on Fog and Edge Computing (ICFEC 2021).
\\
\\Document Version: Preprint Version
\\
\\ \textcopyright 2021 IEEE. Personal use of this material is permitted. Permission from
IEEE must be obtained for all other uses, in any current or future media,
including reprinting/republishing this material for advertising or
promotional purposes, creating new collective works, for resale or
redistribution to servers or lists, or reuse of any copyrighted component of
this work in other works. 
\end{titlepage}

\title{TOD: Transprecise Object Detection to Maximise Real-Time Accuracy on the Edge}

\author{\IEEEauthorblockN{JunKyu Lee, Blesson Varghese, Roger Woods, Hans Vandierendonck}
\IEEEauthorblockA{The Institute of Electronics, Communications and Information Technology, Queen's University Belfast, UK
\\Email: \{junkyu.lee, b.varghese, r.woods, h.vandierendonck\}@qub.ac.uk}}

\maketitle 
\thispagestyle{plain} 
\pagestyle{plain}

\begin{abstract}
Real-time video analytics on the edge is challenging as the computationally constrained resources typically cannot analyse video streams at full fidelity and frame rate, which results in loss of accuracy. This paper proposes a Transprecise Object Detector (TOD) which maximises the real-time object detection accuracy on an edge device by selecting an appropriate Deep Neural Network (DNN) on the fly with negligible computational overhead. TOD makes two key contributions over the state of the art: (1) TOD leverages characteristics of the video stream such as object size and speed of movement to identify networks with high prediction accuracy for the current frames; (2) it selects the best-performing network based on projected accuracy and computational demand using an effective and low-overhead decision mechanism.
Experimental evaluation on a Jetson Nano demonstrates that TOD improves the average object detection precision by $34.7 \%$  over the YOLOv4-tiny-288 model on average over the MOT17Det dataset. In the MOT17-05 test dataset, TOD utilises only $45.1\%$ of GPU resource and $62.7 \%$ of the GPU board power without losing accuracy, compared to YOLOv4-416 model.   
We expect that TOD will maximise the application of edge devices to real-time object detection, since TOD maximises real-time object detection accuracy given edge devices according to dynamic input features without increasing inference latency in practice.    
\end{abstract}
\IEEEpeerreviewmaketitle

\section{Introduction}
\label{sec:introduction}
Edge-based computing looks to leverage the limited computational resources for (pre)processing data using the devices that are located at the boundary of the network.
Computing resources range from small form factor embedded computers available on the user end of the last mile network, right through to complex micro data centres in the infrastructure \cite{varghese-challenges}. 
Performing more processing at the edge acts to improve the responsiveness of applications, as data processed closer to the source reduces the ingress bandwidth demand to the cloud. 

There is a particular benefit to performing video analytics on the edge; it acts to minimise the considerable amount of data that needs to be sent to the cloud. Currently, more powerful Graphical Processing Units (GPUs) such as the NVidia Jetson Nano provide considerable computational capability. In particular, they offer the potential to perform at the edge real-time object detection including both localisation and classification of objects by utilising deep learning-based object detectors. Notice that ``object detection'' in this paper includes both localising and classifying objects, since our paper utilises deep learning-based object detectors rather than machine learning-based object detectors. This ``edge-only mode"~\cite{anan-video} allows a lightweight Deep Neural Network (DNN) resource, thus removing the need for connection to the cloud for such resources. However, this can degrade accuracy compared to a heavyweight object detector if the object sizes are relatively small. In contrast, employing a heavyweight object detector cannot sustain video frame rates in real time, which also degrades accuracy due to the dropped frames. Therefore, it is necessary to select a DNN according to dynamic input frame characteristics. 

Researchers have proposed to select a neural network on the fly according to input image characteristics~\cite{jiang-chameleon, marco-optimizing}. Chameleon~\cite{jiang-chameleon} proposes a runtime decision maker for improving object detection accuracy by employing multiple DNNs and utilising the most suitable of those DNN at any time. However, it does so by periodically re-evaluating the efficacy of the DNNs. This results in significant overhead when the most computationally heavy DNN is used for the periodical evaluation. Moreover, a temporary drop in accuracy can be incurred when attempting DNNs that are not accurate for the current frame. 
Research in \cite{marco-optimizing} proposed to switch between neural networks on the fly with the aim of improving image classification accuracy without the need for the periodic assessment used in \cite{jiang-chameleon}. The lightweight K-Nearest Neighbours (KNN) algorithm was used to select the most suitable DNN out of the variants of MobileNet, Inception Net v2 and ResNet according to input frame features such as average brightness, the level of contrast, and the main object size. This method was developed for image classification, limiting applying it for real-time object detection. For example, a decision maker for real-time object detection needs to be trained relevant to the video streams and the networks and should not be expensive to evaluate at each frame. 

Our goal in this paper is to seek a computationally efficient technique that selects an appropriate DNN on the fly for each video frame in order to improve the real-time detection accuracy (1) by considering characteristics of the video stream that predict the relative accuracy of each network; and (2) by designing an efficient and low-overhead scheduling algorithm that pro-actively decides the best network to use.   

There are many characteristics of video streams~\cite{jiang-chameleon,marco-optimizing} that may be used to predict the object detection precision of different DNNs. In this paper, we explore the insights obtained through the analysis of architectural features of object detection algorithms \cite{korshunov-reducing, huang-speed}. Work in \cite{korshunov-reducing} showed that high frame rates were not necessary in many object tracking applications and found that slower moving objects allowed more frames to be dropped, without affecting the real-time, object tracking accuracy. The trade-off between inference speed and accuracy for object detection algorithms was investigated in  \cite{huang-speed}. The accuracy of lightweight object detectors was found to be equivalent to heavyweight ones for datasets involving relatively large objects. 

The findings of \cite{korshunov-reducing, huang-speed} enable low-overhead and accurate heuristic policies for selecting a DNN on the fly. Indeed, we will demonstrate that it suffices to analyse the sizes of the bounding boxes detected during one frame to predict the best DNN to use in the next frame. Based on this insight, we propose a new  runtime decision maker for edge devices called the Transprecise Object Detector (TOD). It results in a solution that is accurate, but with negligible computational overhead. It moreover improves object detection accuracy by using lightweight DNNs for detecting large, fast-moving objects, and applying heavyweight DNNs for small, slow-moving objects.


The main contributions from this paper are threefold:

\begin{itemize}
    \item A new transprecision technique that selects an appropriate DNN on the fly with negligible computational overhead to maximise real-time object detection accuracy on the edge; the only computational overhead of TOD is in calculating the median of the bounding box sizes per frame, which is negligible compared to the inference latency. 
    \item TOD, a transprecise scheduler that achieves video analytics at real-time speed by pro-active selecting the most appropriate DNN. TOD employs a small number of hyperparameters that can be sought by a grid hyperparameter search; A hyperparameter search returns the optimal hyperparameters for TOD according to the dataset characteristics such as object moving speeds and object sizes and the different average precision and inference speed characteristic from multiple DNNs in order to maximise real-time object detection accuracy on a given computing platform.
    \item Demonstration of the new system improvements in accuracy using the Multiple Object Tracking Challenge 2017 Detection (MOT17Det) datasets. 
\end{itemize}

We present the related work in section~\ref{sec:relatedwork} and follow it by introducing our TOD solution in section~\ref{sec:tod}. The experimental evaluation is presented in section~\ref{sec:experiment} and followed by the discussion in section~\ref{sec:discussion}. The conclusions are presented in section~\ref{sec:conclusion}.  

\section{Related Work} 
\label{sec:relatedwork}
We discuss related work including exploration of effects of dropped frames on accuracy, the methodology to improve the inference throughput, exploitation of resource-accuracy trade-off 
and the methodology to select a DNN on the fly.

In \cite{korshunov-reducing}, a framework was presented that formalises the dependency between dropped frames and algorithms' accuracy. The threshold for the dropped frame rate with accuracy was investigated theoretically for single object detection. It was found to be proportional to the bounding box size divided by the object moving speed. In \cite{mohan-determining}, the authors determined, again theoretically,  the lowest possible frames per second (FPS) against accuracy for object tracking video analytics. 

Improving inference throughput can improve accuracy for real-time object detection, since it can reduce the dropped frames. The crosstalk cascade technique in~\cite{dollar-crosstalk} enables neighbouring detectors to communicate with each other with the aim of accelerating inference without losing accuracy. It achieved a $4$-$30 \times$ speedup over a previous technique which was performed at each image location.
In \cite{yang-rethinking}, an industrial case study on rethinking the design of CNN software was presented and a combination of techniques identified that was shown to improve inference speed.

In \cite{anan-video}, a configurable scheduling was used to provide a resource-accuracy trade-off for video analytics.
The authors in \cite{ali-edge} explored pre-processing on edge devices for large-scale video stream analytics by exploiting a resource-accuracy trade-off for scheduling. The initial pre-processing of the data close to the data source at edge devices minimised the
data size that is transferred and stored in the cloud for an object recognition application.

Transprecise techniques used in \cite{lee-energy, lee-air} adapts precision arithmetic dynamically according to runtime information such as convergence rate and numerical stability of individual computation modules. The transprecise techniques saved energy and accelerated computation without losing accuracy for linear system solvers in \cite{lee-energy, lee-air}.  

In \cite{jiang-chameleon}, a resource manager allocates the computational resource from edge to cloud dynamically to maximise the object detection accuracy given a limited computational resource. It profiles the accuracy of object detection from the current configuration by utilising the inferences from the most heavyweight DNN as the ground truth approximation. The temporal and spatial correlation in the video stream data was exploited to obtain sufficient FPS throughput, but periodical inferences from the most heavyweight DNN were still time-consuming. The resource manager seeks an optimal sufficient FPS for a DNN and uses it to select the most appropriate DNN based on computational resource usage. In \cite{marco-optimizing}, the authors propose switching in on the fly the most suitable neural network, based on the decision from a lightweight KNN classifier to improve image classification accuracy. \cite{marco-optimizing} did not consider input frame features for object detection such as object moving speed, since \cite{marco-optimizing} considered the image classification applications rather than object detection. For real-time objection detection applications, the increased dropped frames due to the inference latency from a lightweight KNN classifier would not be negligible.

\section{TOD: Transprecise Object Detector} 
\label{sec:tod}

\subsection{Motivation: Trade-off in Real-Time Object Detection Accuracy Between Light DNN and Heavy DNN} \label{sec:analysis_rt_accuracy}
\begin{figure}[!t] 
\centering
\includegraphics[width=\linewidth]{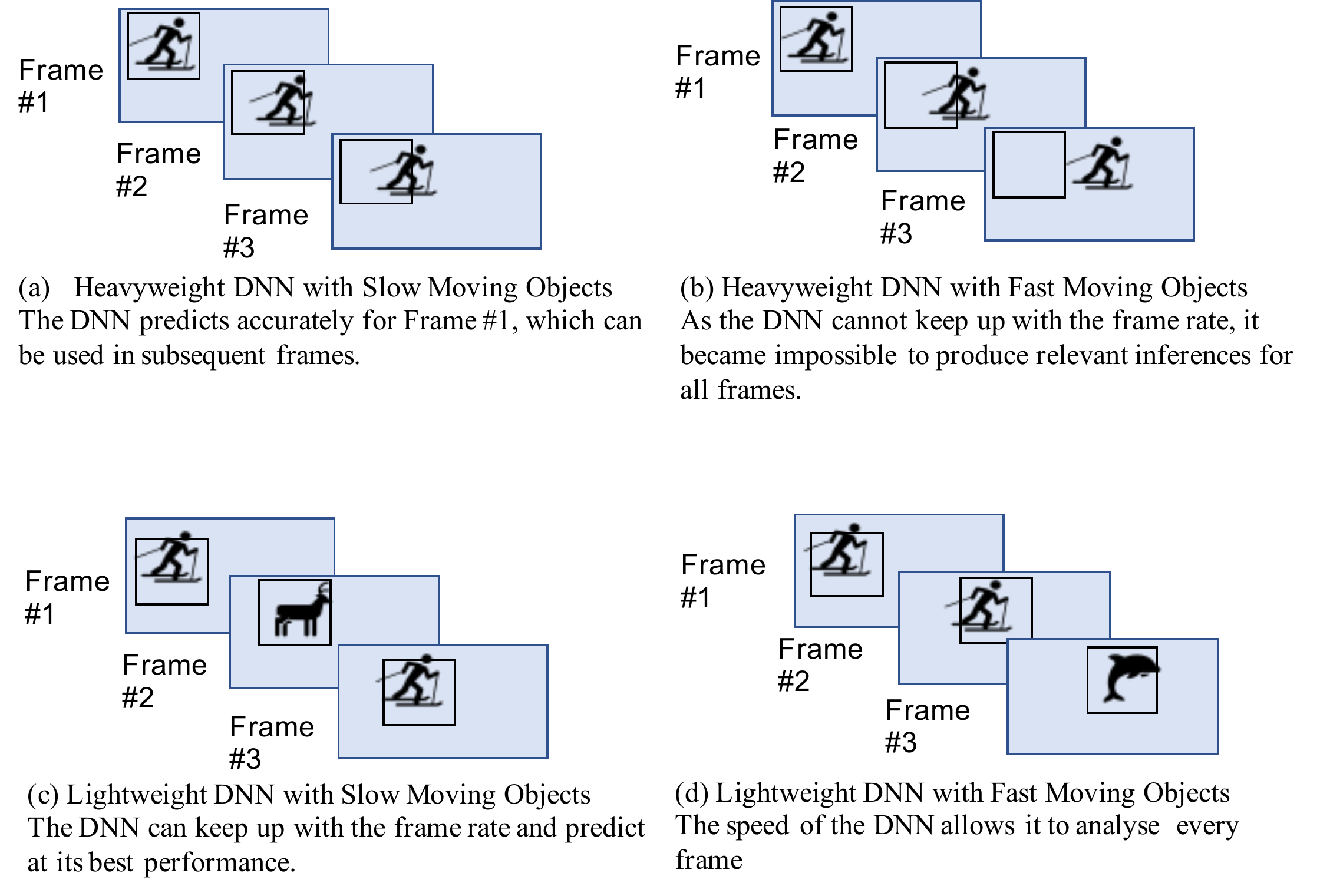}
\caption{DNN Inference Characteristics}
\label{fig:moving_speed_combined}
\end{figure}

The optimal DNN that produces the best accuracy for real-time object detection depends on object moving speed, object size and computational capabilities. For example, the speed of a moving object  can be a critical factor in determining the permissible dropped frame rate \cite{korshunov-reducing}. 
Fig.~\ref{fig:moving_speed_combined} represents the comparison in terms of accurate inference between detecting an object moving slowly and quickly using a heavyweight DNN and a lightweight DNN.  Accurate detection of an object with a bounding box in Frame \#1 is achieved in Fig.~\ref{fig:moving_speed_combined}.(a) and (b) due to the inference using an accurate DNN. However, if the DNN inference time does not meet the real-time threshold as demonstrated in Fig.~\ref{fig:moving_speed_combined}.(b), dropped frames will occur for Frame \#2 and \#3 as the bounding box location does not change accurately enough to reflect the movement. 
The bounding box location information predicted at Frame \#1 could be used for Frame \#2 when the object moves slowly as demonstrated on (a) in Fig.~\ref{fig:moving_speed_combined}, while it is not applicable when the object moves fast as on (b).

Fig.~\ref{fig:moving_speed_combined}.(c) and (d) represent the inferences when using a lightweight DNN for object detection.
It will catch every frame in real-time as demonstrated by bounding box location located around the object, but the inference has limited accuracy and might result in misclassification of the object. 
%

A heavyweight DNN is more accurate for inference for each individual frame. However, inference from a heavyweight DNN on resource contrained device is slow and cannot be performed at the real-time frame rate. As such, the heavyweight DNN is stuck analysing an old frame while the video content in subsequent frames is changing rapidly. Therefore, a heavyweight DNN generates more dropped frames. 

\begin{figure}[!t] 
 \centering
 \includegraphics[width=3.0in]{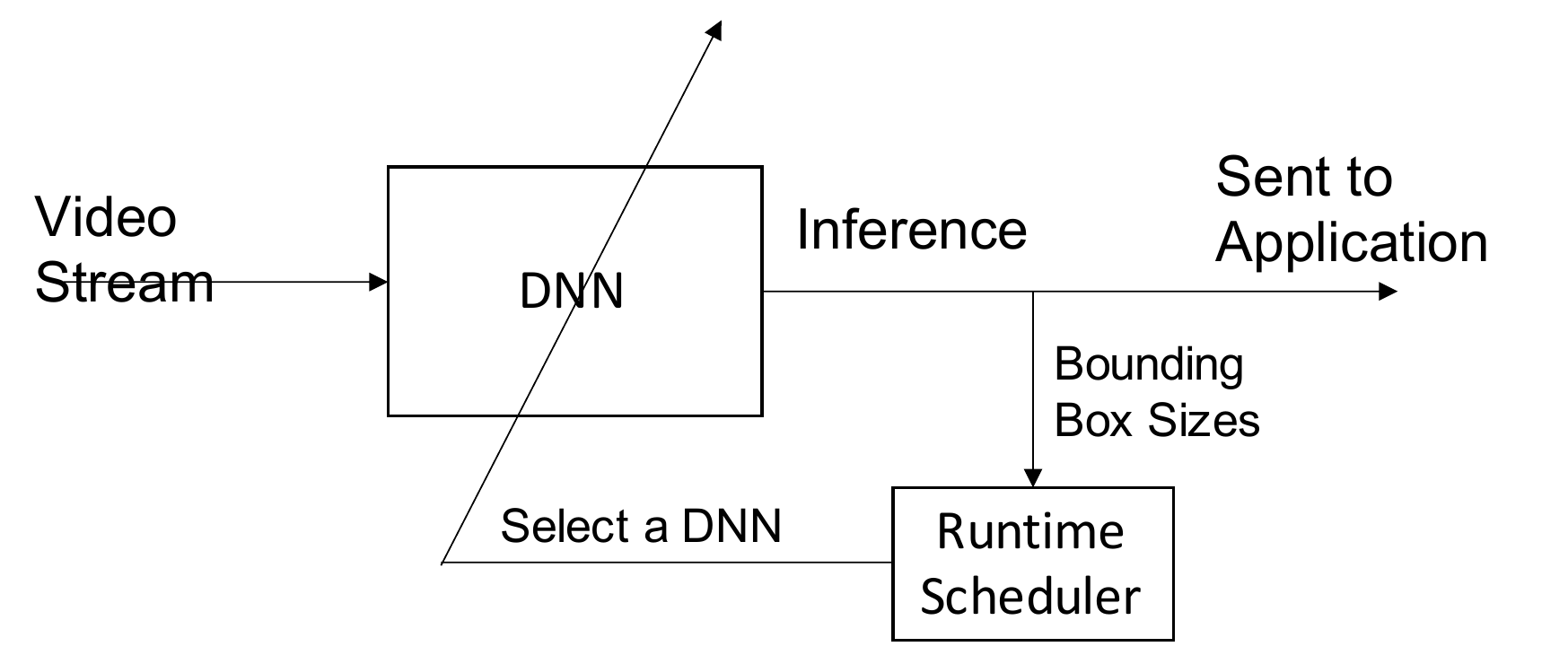}
 \caption{TOD Architecture Overview}
 \label{fig:tod_architecture}
 \end{figure}

Thus, a trade-off exists between the accuracy for real-time object detection applications and dropped frames. Using this observation, the TOD is able to exploit this trades-off for real-time object detection by utilising the different precision characteristics of each DNN. Fig.~\ref{fig:tod_architecture} describes the TOD architecture overview. 

\subsection{Runtime Scheduler Implementation} 
\label{subsec:runtime scheduler}

\subsubsection{TensorRT pre-trained YOLO}
We utilise four pre-trained YOLO inference models trained on the COCO dataset in~\cite{jkjung}. Four inference modules named as YOLOv4-tiny-288, YOLOv4-tiny-416, YOLOv4-288 and YOLOv4-416 for the tiny YOLO version with $288$ and $416$ input sizes and full YOLO version having $288$ and $416$ input sizes, respectively. The modules have been optimised for the Jetson Nano platforms using the NVidia TensorRT optimisation tool. The Python API implementation for our TOD inference is as follows: 
\begin{equation} \label{eq: impl_API}
boxes, scores, classes = detect\_objects(DNN, frame), 
\end{equation}
where the \emph{detect\_objects} API function takes two arguments, \emph{DNN} and \emph{frame}. The \emph{boxes} contain all detected bounding boxes information in the \emph{frame}, the \emph{scores} contain the confidence score for each bounding box detected, and the classes contain the label information for an object for each bounding box detected. Bounding boxes with a confidence score larger than $0.35$ and classes that are assigned `person', are considered. 

With four YOLO DNNs \emph{preloaded} before initiating object detection tasks, our runtime scheduler (Algorithm~\ref{alg:runtime scheduler}) then switches a neural network with no time overhead in practice. In other words, it just requires to switch a pointer location to a DNN stored in memory in the Jetson Nano. The \emph{DNN} in (\ref{eq: impl_API}) is determined from the previous frame inference. We choose YOLOv4-416 for the default option.  

\subsubsection{Implementation of Inference given Fixed FPS for Real-Time Accuracy Measurement}
\begin{figure}[!t] 
\centering
\includegraphics[width=3.0in]{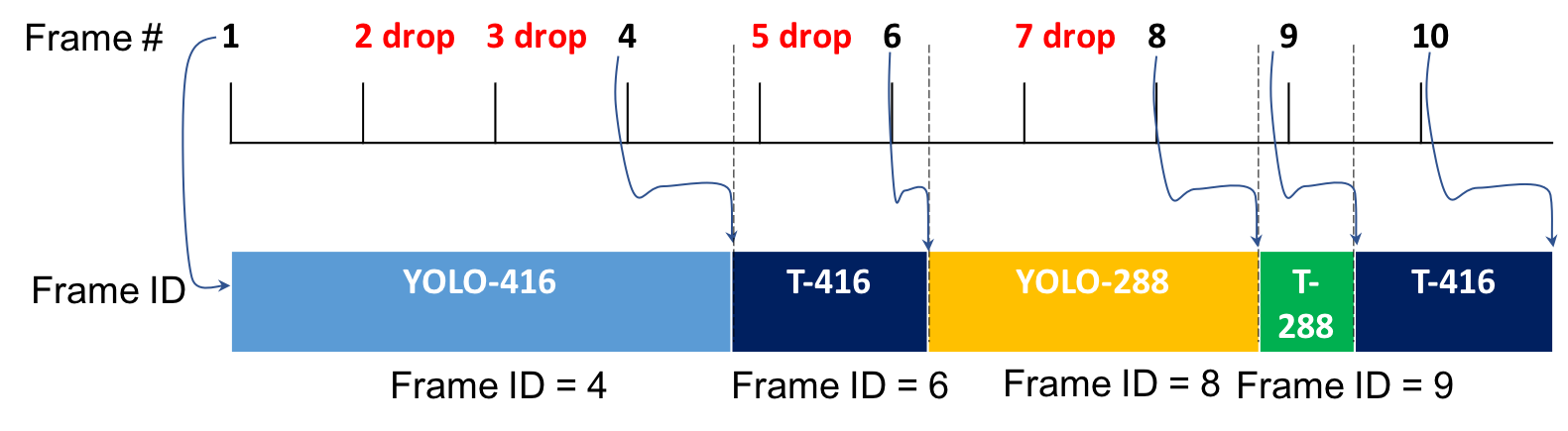}
\caption{Implementation of Fixed FPS Mechanism for Accuracy Measurement}
\label{fig:fixed_fps_impl}
\end{figure}

Fig.~\ref{fig:fixed_fps_impl} outlines details of how the inference framework operates, given an FPS constraint.  
We utilise the location information detected from the previous frame for the accuracy measurement for the dropped frames, as used in \cite{jiang-chameleon}. For example, YOLOv4-416 is initially chosen in Fig.~\ref{fig:fixed_fps_impl} and the location information for the first frame, i.e., Frame \#1 will replace the prediction results for the dropped Frame \#2 and \#3. When YOLOv4-tiny-416 is chosen, Frame \#4 is available and the inference comes from it, and so on. This can be implemented with the drop frame option in the Gstreamer appsink for real-time applications.  The implementation of the algorithm for Fig.~\ref{fig:fixed_fps_impl} is described in Algorithm~\ref{alg:pseudo-droppedframes}.     

\begin{algorithm}
\begin{algorithmic}
\STATE \textbf{Set Hyperparameters}: $H_{opt}$ = \{$h_1^*$, $h_2^*$, $h_3^*$\}
\\ \textbf{ }
\STATE \textbf{Initialization}: Obtain the four YOLOs and load them into memory
\\$DNN_1$ = YOLOv4-tiny-288
\\$DNN_2$ = YOLOv4-tiny-416
\\$DNN_3$ = YOLOv4-288
\\$DNN_4$ = YOLOv4-416
\\Load($DNN_1$, $DNN_2$, $DNN_3$, $DNN_4$) to Memory 
\\$median(bboxes)_0$ = 0
\\ \textbf{ }
\\ \textbf{Runtime Selection}:
\FOR{t = $1, 2, \dots$}
\STATE Get an image sample $img_t$
\STATE Compute $median(bboxes)_{t-1}$ detected from $img_{t-1}$.
\IF {$median(bbox)_{t-1} > h_3^*$}
        \STATE $DNN = DNN_1$ 
\ELSIF {$h_2^* < median(bboxes)_{t-1} \leq h_3^*$}
\STATE $DNN = DNN_2$ 
\ELSIF {$h_1^* < median(bboxes)_{t-1} \leq h_2^*$}
\STATE $DNN = DNN_3$ 
	\ELSE   
\STATE $DNN = DNN_4$ 	
	\ENDIF  
\STATE boxes, scores, classes = detect\_objects($DNN$, $img_t$) 
\ENDFOR
\end{algorithmic}
\caption{Runtime Scheduler Algorithm}
\label{alg:runtime scheduler}
\end{algorithm}

\begin{algorithm}
\begin{algorithmic}
\STATE {\textbf{When DNN inference time is longer than 1/FPS}}
\IF {FrameID $>$ Frame\#} 
    \STATE \textbf{Use Previous Inference for Current Inference}  
    \\boxes = pre-boxes \\classes = pre-classes \\scores = pre-scores
\ELSE   
    \STATE \textbf{Update Accumulated Inference Time and FrameID}
    \\acc\_inf\_time = acc\_inf\_time + dnn\_time
    \\FrameID = int(acc\_inf\_time$\times$FPS) + 1
\ENDIF
\STATE{ }
\STATE{\textbf{When DNN inference time is shorter than 1/FPS}}
\IF {acc\_inf\_time $<$ (Frame\#/FPS)} 
    \STATE{acc\_inf\_time $=$ (Frame\#/FPS)}
\ENDIF
\end{algorithmic}
\caption{Pseudocode for Predictions for Dropped Frames}
\label{alg:pseudo-droppedframes}
\end{algorithm}

\subsubsection{Runtime Scheduler}
By leveraging the findings from \cite{korshunov-reducing} and \cite{huang-speed}, our runtime scheduler controls the deployment frequency of each DNN to maximise the accuracy based on the object sizes detected. Exploiting the research findings from \cite{huang-speed}, namely that the accuracy of a lightweight DNN is equivalent to its heavyweight counterpart for large object detection, allows our runtime scheduler to select an appropriate DNN for the next frame, based on the object sizes. To implement an adequate runtime policy, we consider the Median of Bounding Box Sizes (MBBSs) as a representative parameter to control the deployment frequency of each DNN according to the dataset. We choose MBBS as a representative parameter since (i) object detection accuracy from a DNN mainly depends on object sizes, (ii) a lightweight DNN can produce equivalent accuracy to a heavyweight DNN for larger objects \cite{huang-speed} and (iii) the median of the object sizes can be more reliable than the average of the object sizes against false positives (e.g., sometimes, entire frames were detected as false positives.).  

Our runtime scheduler utilises $(n_{DNN}-1)$ hyperparameters to assign a inference task to one out of $n_{DNN}$ DNNs. For example, TOD runtime scheduler has the runtime policy: 
\begin{itemize}
    \item $0\% < MBBS \leq h_1\%$: YOLOv4-416 (the heaviest-weight DNN);
    \item $h_1\% < MBBS \leq h_2\%$: YOLOv4-218; 
    \item $h_2\% < MBBS \leq h_3\%$: YOLOv4-tiny-416;
    \item $h_3\% < MBBS$: YOLOv4-tiny-288 (the lightest-weight DNN),
\end{itemize} 

where $h_1$, $h_2$ and $h_3$ are hyperparameters such that $h_1 < h_2 < h_3$ and $h_1$ means that the median of the bounding box sizes, e.g., height $\times$ width, in a frame occupies $h_1\%$ of the image. In the policy, YOLOv4-218 is used for the next frame if the median of the bounding box sizes is between $h_1$ and $h_2\%$.

The next question is how to consider the object moving speed in this runtime policy framework. For example, the runtime scheduler should provide heavyweight DNNs with more opportunities for detecting objects moving slowly based on Fig.~\ref{fig:moving_speed_combined}.(a) and (b). In contrast, the runtime scheduler should provide lightweight DNNs with more opportunities for detecting objects moving faster based on Fig.~\ref{fig:moving_speed_combined}.(c) and (d). We employ the hyperparameter search approach to consider the object moving speed. For example, if objects moves fast, e.g., when cars are on the highway, the optimal set by the hyperparameter search would increase the region for YOLOv4-tiny-288 by reducing $h_3$, since it would improve the accuracy. The hyperparameter search considers object moving speed to select the optimal hyperparameter set, $\{h_1, h_2, h_3\}$.

\subsubsection{Hyperparameter Search}
Our implementation on an NVidia Jetson Nano has $n_{DNN} = 4$, implying three hyperparameters required, $h_1$, $h_2$, and $h_3$. We examined eight different hyperparameter sets, $H^{(i,j,k)}$, where $H^{(i,j,k)} = \{ h_1^{(i)} = \{0.0007, 0.007\}, h_2^{(j)} = \{0.008, 0.03\}, h_3^{(k)} =\{0.04, 0.1\} \}$, where $i,j,k = \{1, 2\}$. 
The optimal hyperparameter set, $H_{opt} = \{0.007, 0.03, 0.04\}$ shows the best average precision accuracy over MOT17Det datasets having 30 FPS constraints. 

Table~\ref{tbl:table-hyparam} shows our hyperparameter search results. We choose $\{0.007, 0.03, 0.04\}$ for $H_{opt}$ over $\{0.007, 0.03, 0.1\}$, since $H_{opt}$ can utilise the most lightweight DNN more often. The ground truth information is provided with MOT17Det training set. For example, the ground truth has nine column's data as follows: $ 1, -1, 794.2, 47.5, 71.2, 174.8, 1, class ID, 0.8 $ \cite{MOT17Det}. The columns represent the frame number, -1 (detection), the bounding box left side pixel coordinate, the bounding box top side pixel coordinate, the bounding box width, the bounding box height, the confidence score, the class ID, and the visibility respectively. 
For detection applications, the visibility information is not meaningful, so we intentionally put `-1' in the field for the inference data generated from our TOD.     
\begin{table}
  \caption{Hyperparameter Search}
  \label{tbl:table-hyparam}
  \centering
  \includegraphics[width=\linewidth]{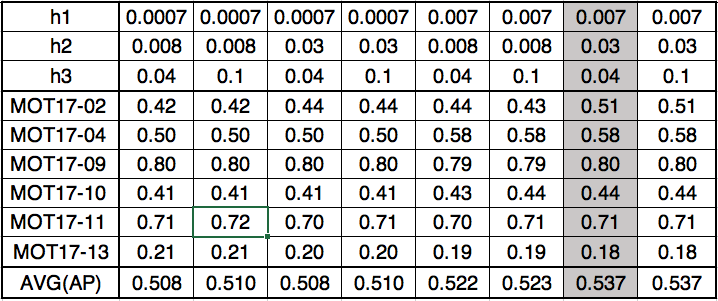}
\end{table}

There are three video data generated by a static camera, MOT17-02, 04 and 10, two video data generated by a camera moving with a walking speed, MOT17-09 and 11, and one video data generated by a camera moving with a car speed. This explains the average precision characteristic similarity between MOT17-02, 04 and 10  (as YOLOv4-416 is the best DNN for the three cases), and between MOT17-09 and 11 (as YOLOv4-tiny-416 is the best DNN for the two cases), as illustrated in Fig.~\ref{fig:accuracy_realtime}. The average precision plot in MOT17-13 is significantly different to others; YOLOv4-288 shows the best performance since the small objects are moving faster compared to other datasets. Therefore, the best DNN depends on the situation, and our hyperparameter search seeks an optimal hyperparameter set for the average case between the three different cases so that TOD can be applied to any of the three cases while keeping the best performance.  

\section{Experimental Evaluation} \label{sec:experiment}
\subsection{Experimental Setting}
The resources used are as follows:
\\- Computing Platform: An NVidia Jetson Nano Board (MAX power mode). 
\\- Object Detectors: Four YOLO version 4 algorithms optimised by TensorRT with FP16 (e.g., half precision) option, including YOLOv4-288, YOLOv4-416, YOLOv4-tiny-288 and YOLOv4-tiny-416, in \cite{jkjung}. 
\\- Data: MOT17Det datasets \cite{leal-motchallenge15, milan-mot16, MOT17Det}. 
\\- Accuracy Evaluation Tool: Matlab interface MOT evaluation tool kit provided by \cite{MOT17Det}. We pre-processed ground truth data from MOT17Det by converting the flags (confidence scores) from 1 to 0 if the label is neither pedestrian nor static person to ignore non-person label data from ground truth for the accuracy evaluation \cite{MOT17Det}.
\\- Profiling Tool: NVidia Tegrastats tool with 1 second resolution (Tegrastats Default Option).  

\subsection{Accuracy Evaluation} 
We firstly demonstrate the offline mode accuracy followed by the real-time mode accuracy.
\subsubsection{Offline Mode}
\begin{figure}[!t] 
\centering
\includegraphics[width=3.2in]{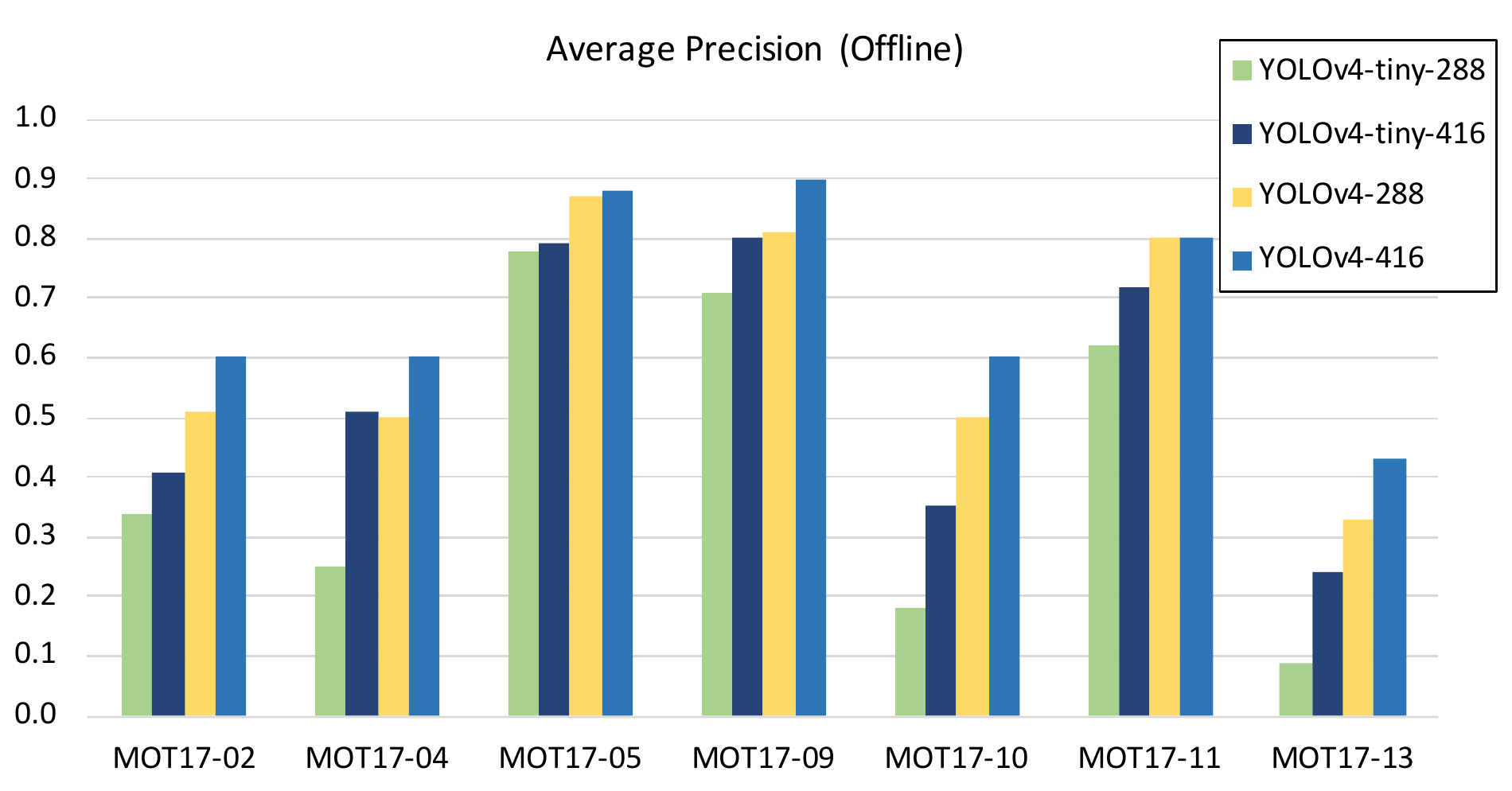}
\caption{Average Precision (Offline Mode)}
\label{fig:accuracy_offline}
\end{figure}

\begin{figure}[!t] 
\centering
\includegraphics[width=3.2in]{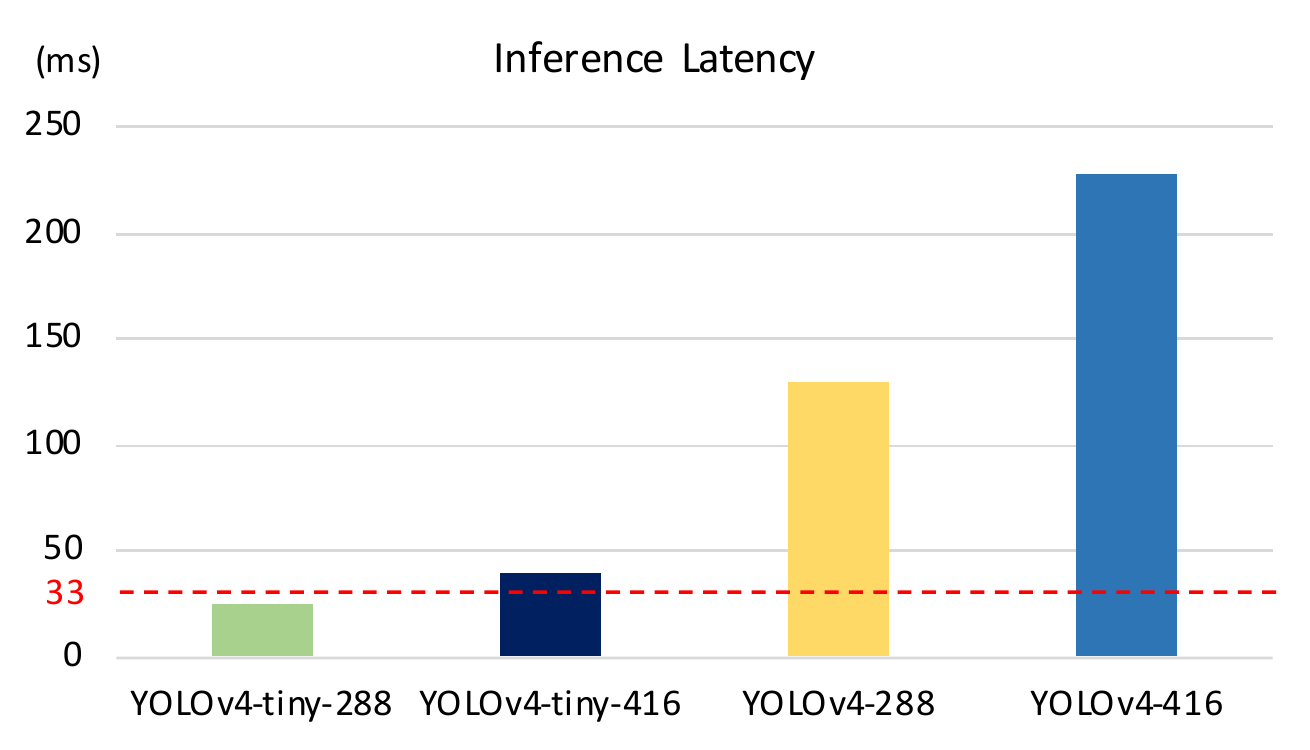}
\caption{Inference Latency}
\label{fig:latency_offline}
\end{figure}

In the offline mode, accuracy is evaluated on all frames, i.e., no dropped frames for DNNs. Fig.~\ref{fig:accuracy_offline} represents the average precision from four YOLO algorithms on the MOT17Det datasets. The most heavyweight DNN, namely YOLOv4-416, shows the best accuracy for all datasets and the most lightweight DNN, namely YOLOv4-tiny-288, shows the worst accuracy for all datasets. The average inference latency for each DNN on the Jetson Nano is described in Fig.~\ref{fig:latency_offline}. Only YOLOv4-tiny-288 meets the threshold for inference latency for real-time given a 30 FPS constraint. Other DNNs will drop frames due to a longer inference latency than the real-time threshold ($1/30$ second).

\subsubsection{Real-Time Mode}
\begin{figure}[!t] 
\centering
\includegraphics[width=3.2in]{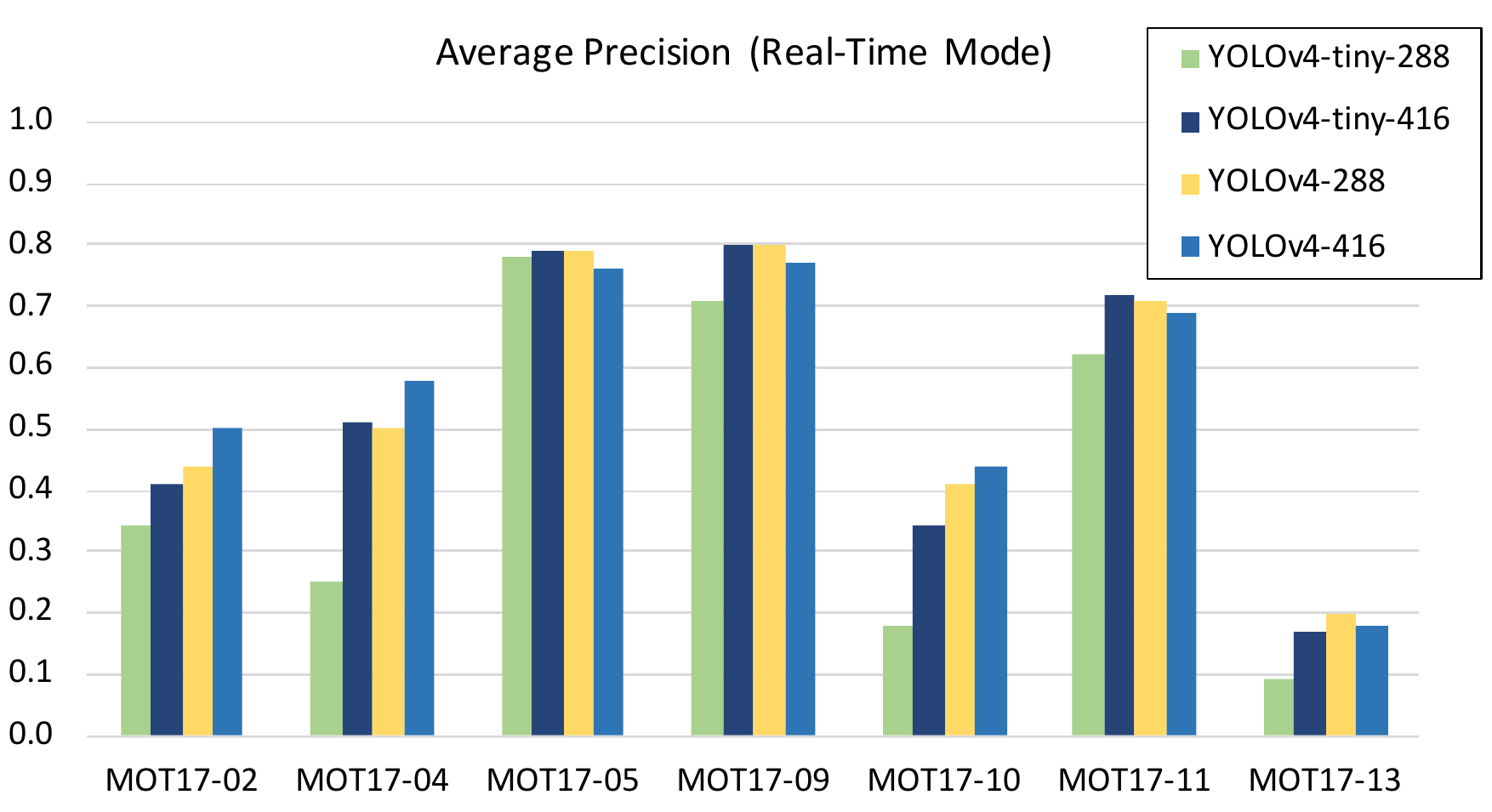}
\caption{Average Precision (Real-Time Mode)}
\label{fig:accuracy_realtime}
\end{figure}

\begin{figure}[!t] 
\centering
\includegraphics[width=2.8in]{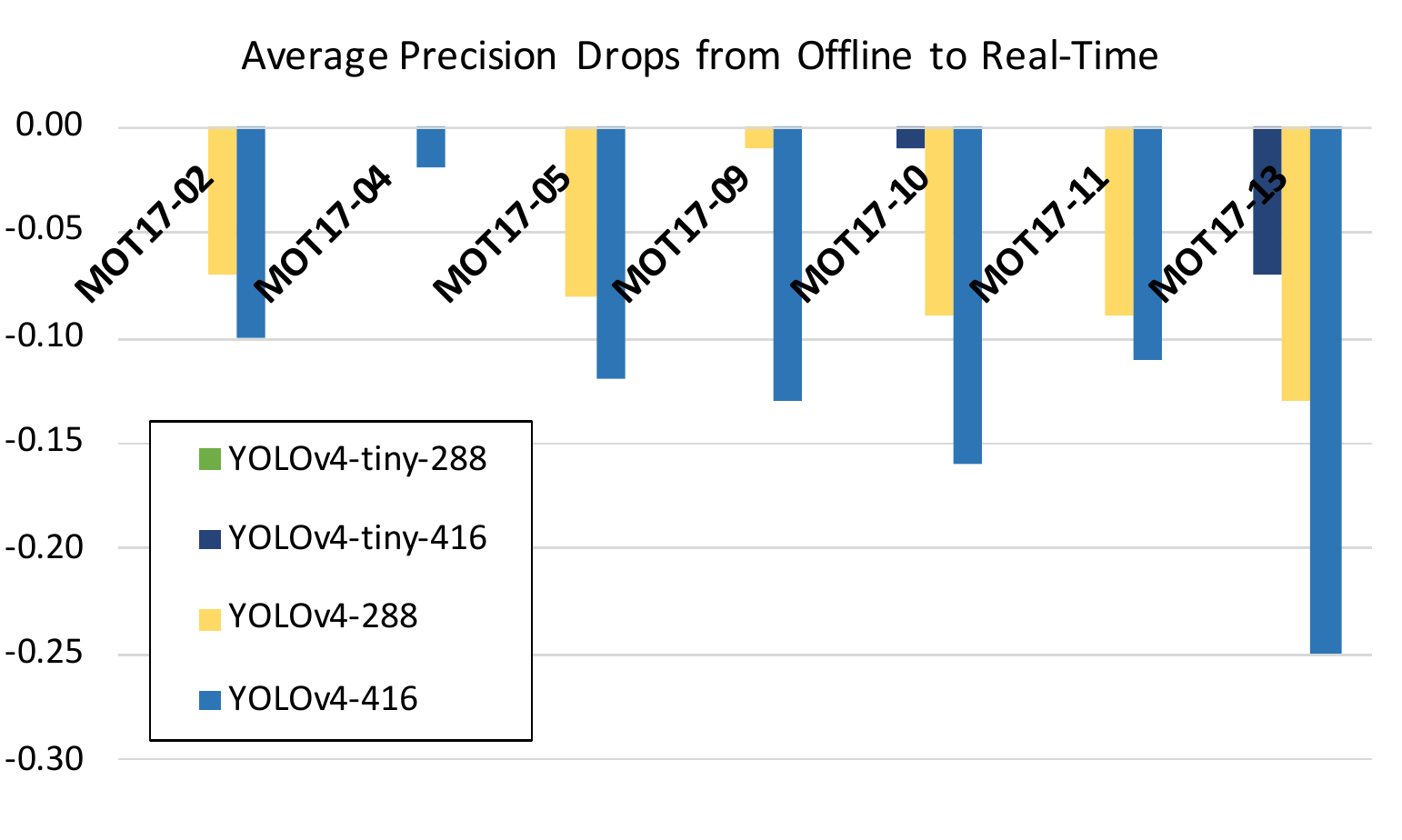}
\caption{Average Precision Drop from Offline to Real-Time}
\label{fig:accuracy_drop}
\end{figure}

Due to dropped frames by a FPS constraint, the average precision for the real-time mode is different to the offline mode. For the real-time mode accuracy measurement, Algorithm~\ref{alg:pseudo-droppedframes} is used. Fig.~\ref{fig:accuracy_realtime} shows the average precision for the real-time mode with MOT17-02, 04, 09, 10, 11, and 13 datasets with a 30 FPS constraint and the MOT17-05 dataset with a 14 FPS constraint. 
The accuracy from the YOLOv4-tiny-288 is unchanged, since it can process every frame in real-time, while the accuracy from the YOLOv4-416 significantly drops due to dropped frames. Fig.~\ref{fig:accuracy_drop} shows the accuracy drop for each DNN on different datasets. The accuracy drop is significant for the MOT17-13 dataset (refer to \url{https://motchallenge.net/vis/MOT17-13}) in which pedestrians move fast, due to a fast moving camera. In contrast, the accuracy drop is not significant for the MOT17-04 dataset (refer to \url{https://motchallenge.net/vis/MOT17-04}) as pedestrians move slower due to a static camera installed at a distance.  

\subsubsection{Real-Time Accuracy from TOD}
\begin{figure}[!t] 
\centering
\includegraphics[width=3.2in]{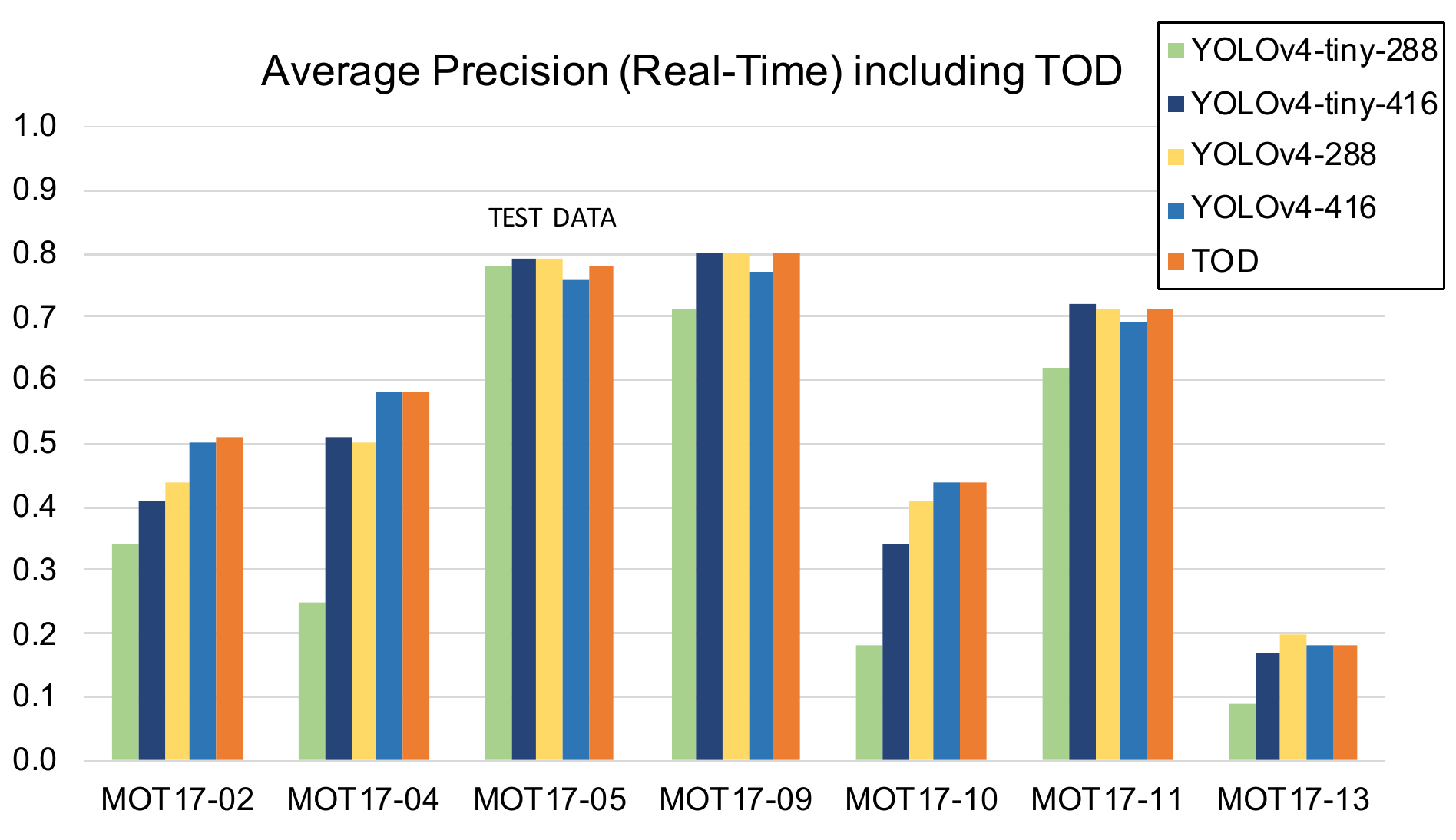}
\caption{Average Precision Comparison}
\label{fig:accuracy_hybrid}
\end{figure}

Fig.~\ref{fig:accuracy_hybrid} shows the average precision comparison of TOD to the four individual DNNs. \emph{Our 
TOD keeps the equivalent average precision to the best average precision for entire dataset.}  The hyperparameter set is chosen based on the datasets having a 30 FPS constraint, and our hybrid DNN TOD is tested with the MOT17-05 dataset having a 14 FPS constraint. TOD shows 0.78 for average precision with the MOT17-05 dataset; the best accuracy was 0.79 by YOLOv4-tiny-416. Overall, TOD shows either equivalent or better accuracy for the four datasets, while minor accuracy loss for the three datasets (e.g., 0.1 average precision loss for MOT17-05 and MOT17-11, and 0.2 loss for MOT17-13.). TOD improves the average precision by $34.7, 7.0, 3.9, 2.0 \%$, compared to YOLOv4-tiny-288, YOLOv4-tiny-416, YOLOv4-288 and YOLOv4-416 respectively on average over entire MOT17Det datasets. 

\subsection{Bounding Box Sizes and Deployment Frequency of DNN}
\begin{figure}[!t] 
\centering
\includegraphics[width=2.8in]{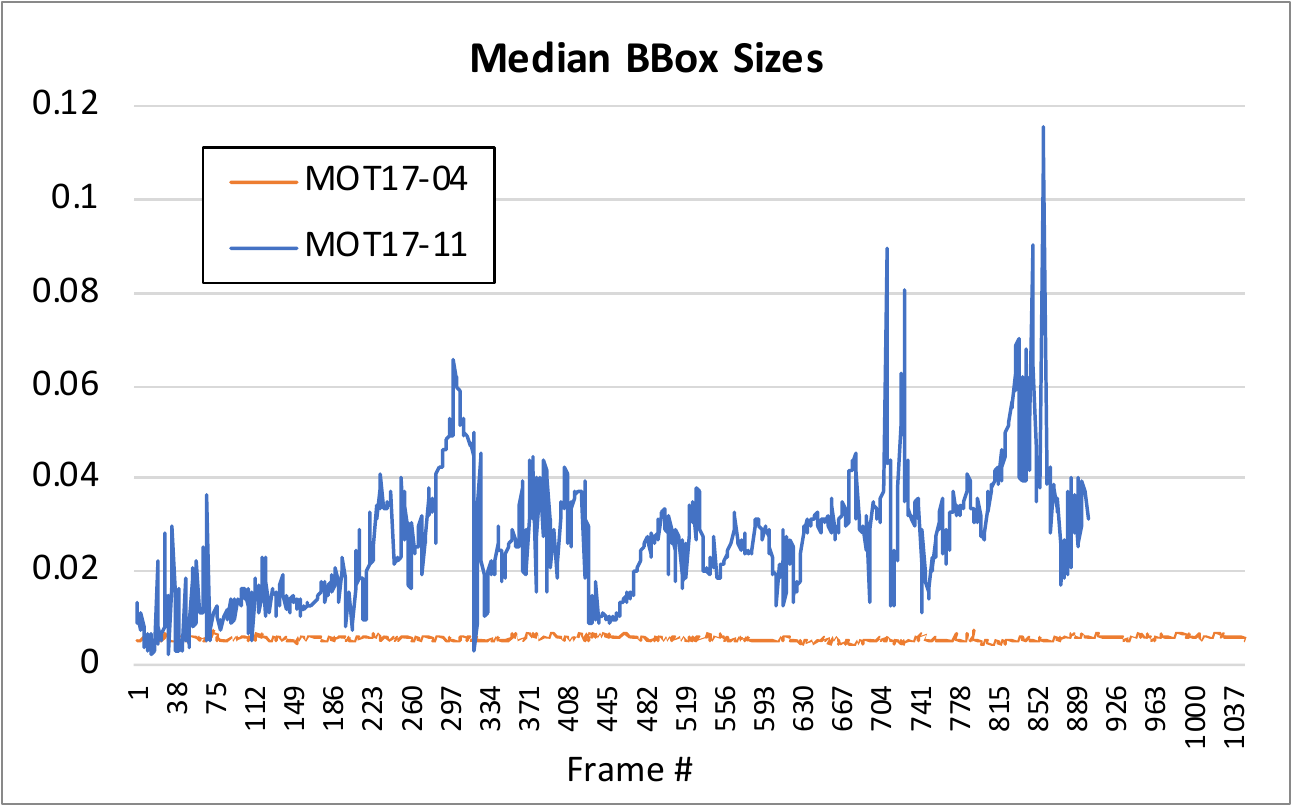}
\caption{Medians of Bounding Box Sizes}
\label{fig:bbox_sizes}
\end{figure}

Fig.~\ref{fig:bbox_sizes} describes the medians of the object sizes (i.e., the bounding box sizes, $width \times height$) for the MOT17-04 and MOT17-11 datasets.       
The variance of bounding box sizes from MOT17-04 generated by a static camera is low, whereas from the MOT17-11 generated by a moving camera is high. TOD stays with YOLOv4-416 for the MOT17-04 dataset, while selecting a DNN from all YOLO variants for the MOT17-11 dataset based on $H_{opt}$ chosen by the hyperparameter search. 

\begin{figure}[!t] 
\centering
\includegraphics[width=\linewidth]{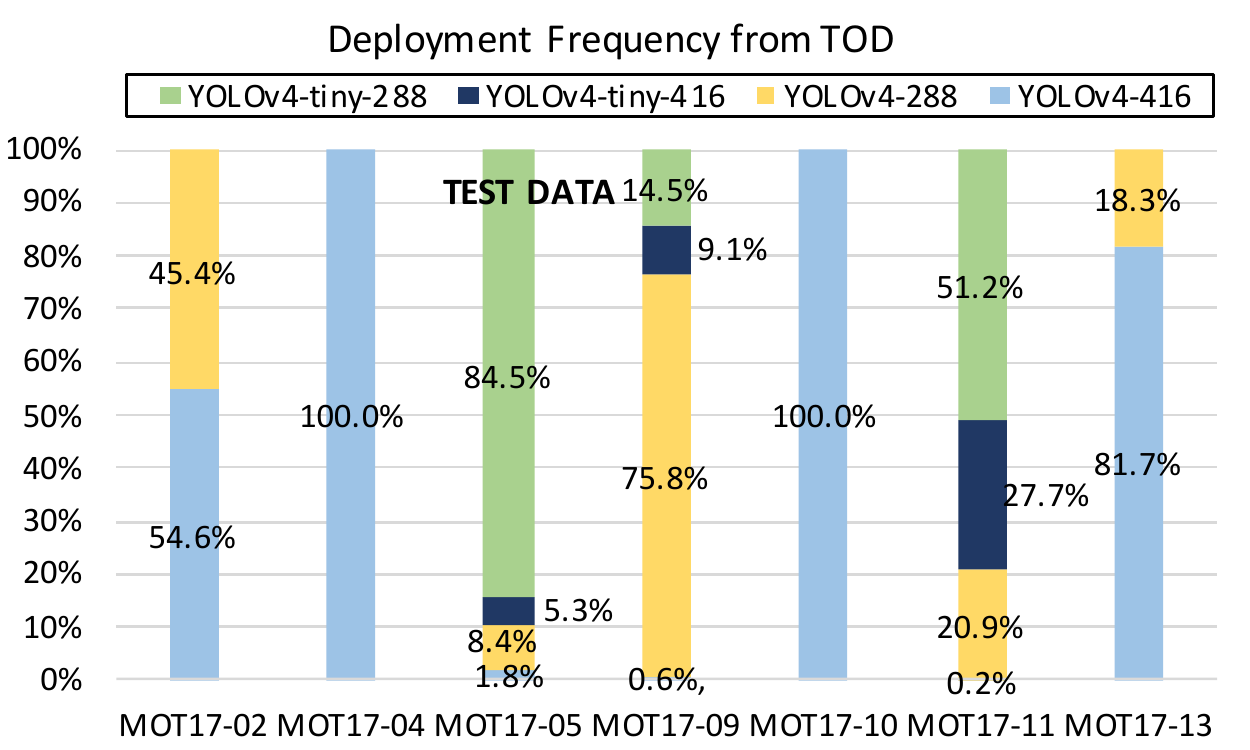}
\caption{Deployment Frequency of Each Network by TOD}
\label{fig:frequency_tod}
\end{figure}

Fig.~\ref{fig:frequency_tod} describes the deployment frequency of each DNN in TOD. In the MOT17-11 dataset, TOD improves the accuracy significantly compared to YOLOv4-tiny-288, while mostly utilising YOLOv4-tiny-288. In the MOT17-05 test dataset, TOD utilises YOLOv4-tiny-288 dominantly with 84.5\%, since YOLOv4-tiny-288 is sufficient for the dataset. 

\subsection{Memory/GPU Utilisation and Power Consumption}
\begin{figure}[!t] 
\centering
\includegraphics[width=3.0in]{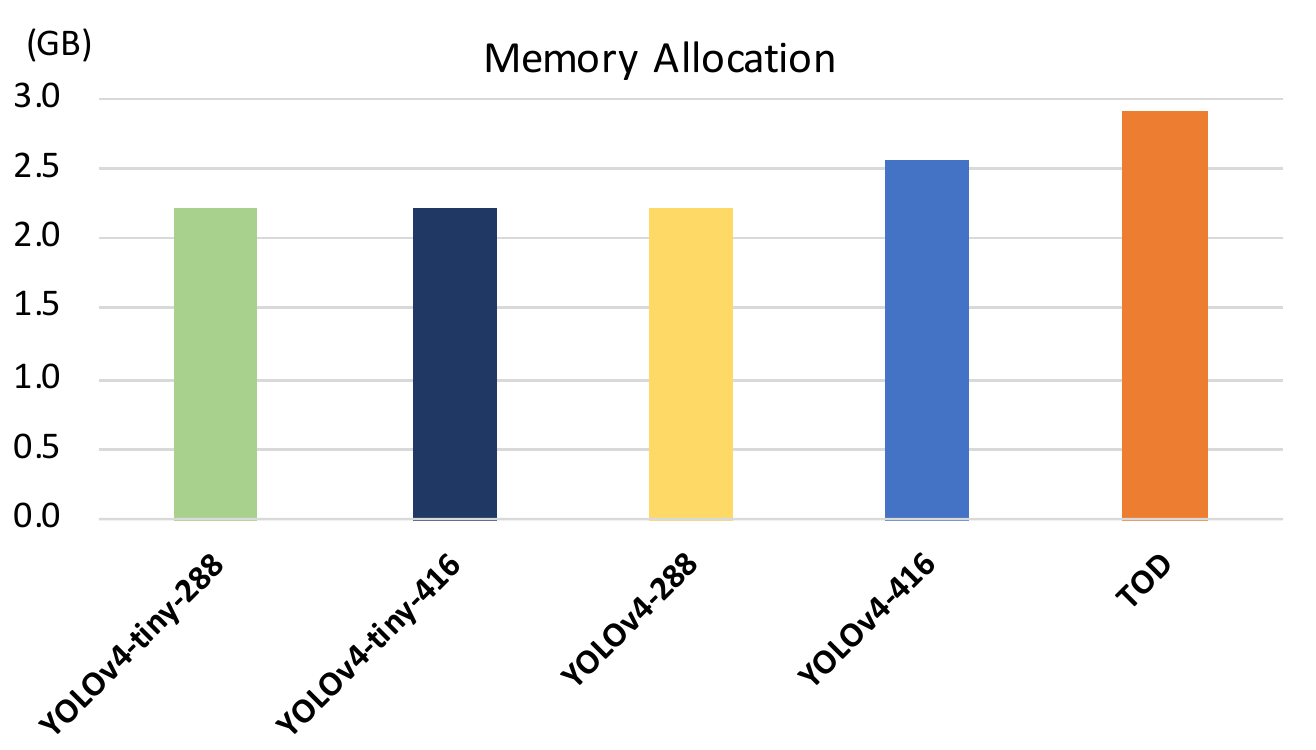}
\caption{Memory Allocation on Jetson Nano with MOT17-05}
\label{fig:mem_alloc}
\end{figure}

Fig.~\ref{fig:mem_alloc} describes the memory allocation according to each DNN when we run each DNN with MOT17-05 dataset; it was implemented with a Gstreamer drop frame option, resulting in the equivalent execution times from different YOLO models.   
TOD loads the four DNNs and requires $\sim$11\% of more memory allocation compared to single YOLOv4-416. The memory has been allocated by 2.21, 2.21, 2.22, 2.56, and 2.85 GB for YOLOv4-tiny-288, YOLOv4-tiny-416, YOLOv4-288, YOLOv4-416 and TOD respectively. Therefore, the memory allocation by TOD is comparable to utilising single YOLOv4-416 on the Jetson Nano. Before loading any DNNs, 1.5 GB are allocated initially. 

\begin{figure}[!t] 
\centering
\includegraphics[width=3.0in]{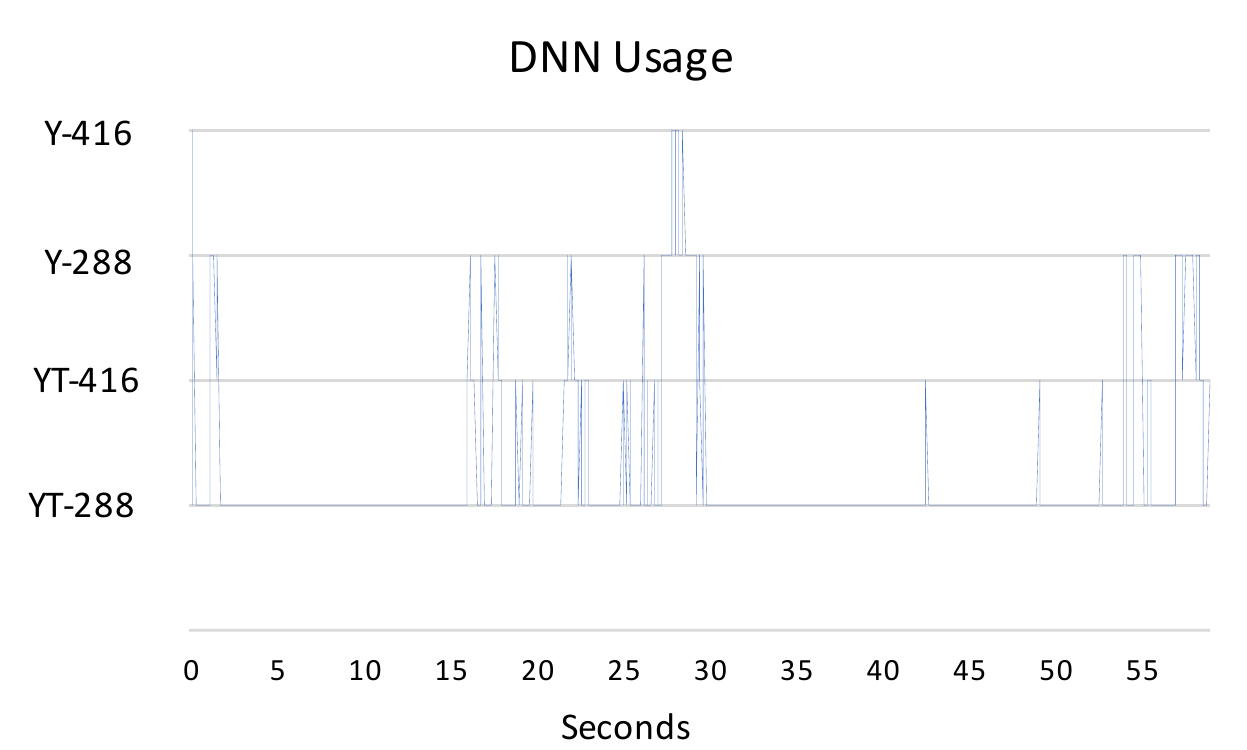}
\caption{DNN Usage of TOD with MOT17-05}
\label{fig:dnn_usage}
\end{figure}

Fig.~\ref{fig:dnn_usage} shows the DNNs TOD selects for the MOT17-05 dataset. In the figure, YT-288, YT-416, Y-288 and Y-416 represent YOLOv4-tiny-288, YOLOv4-tiny-416, YOLOv4-288, and YOLOv4-416 respectively. TOD mostly selects YT-288.  

\begin{figure}[!t] 
\centering
\includegraphics[width=2.8in]{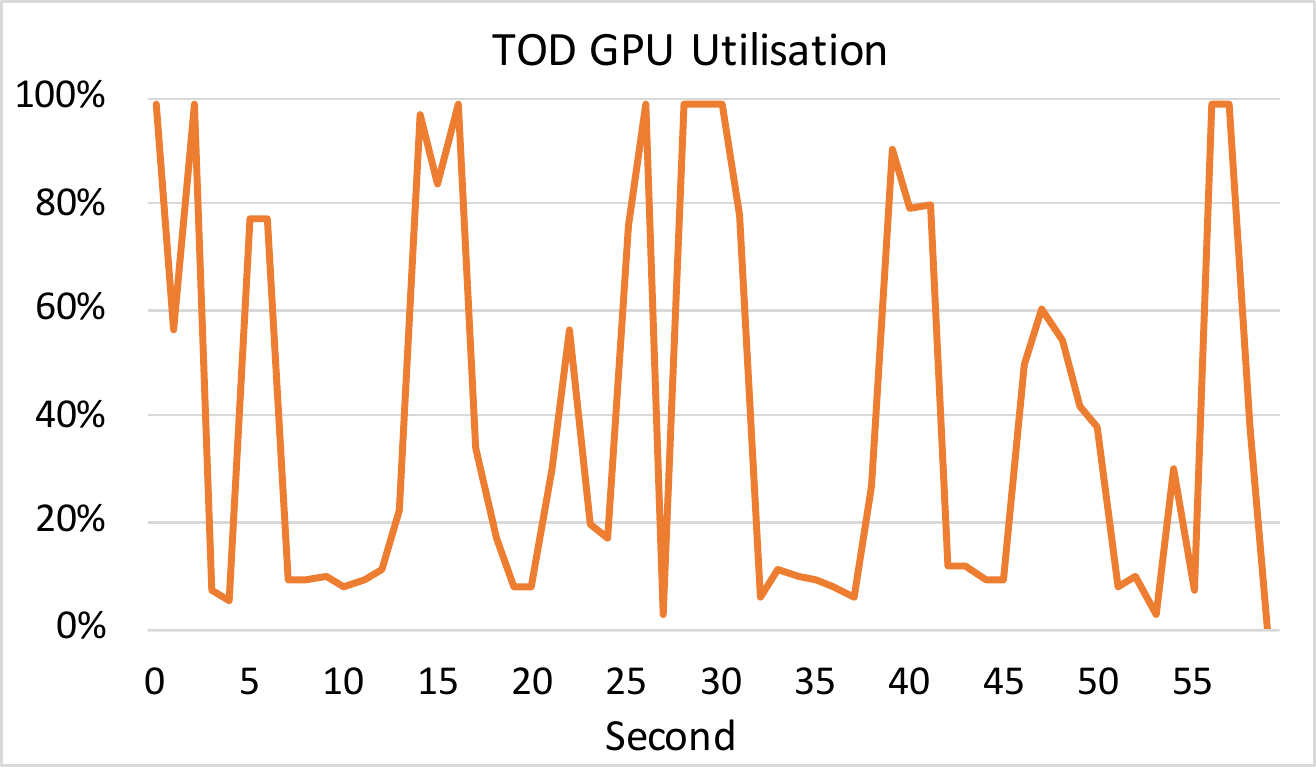}
\caption{GPU Utilisation on Jetson Nano with TOD using MOT17-05}
\label{fig:gpu_util_tod}
\end{figure}

Fig.~\ref{fig:gpu_util_tod} shows the GPU core utilisation for TOD using MOT17-05 dataset; GPU utilisation is the percentage of the GPU engine that is used each clock cycle. TOD utilises 41.1\% of GPU cores on average to run TOD after loading the four YOLOs. Between 15 and 30 seconds, TOD utilises a relatively higher proportion of GPU cores (53.0 \% on average) due to the execution partially from full YOLO models as shown in Fig.~\ref{fig:dnn_usage}. We observed that 84 and 91 \% of GPU cores were used on average to run YOLOv4-288 and YOLOv4-416 after loading the DNN to the memory respectively. \emph{TOD utilises less GPU resource than YOLOv4-416 while maintaining the full YOLO's accuracy in Fig.~\ref{fig:accuracy_hybrid}: only $45.1\%$ of GPU resource is utilised without losing accuracy, compared to YOLOv4-416.}   

\begin{figure}[!t] 
\centering
\includegraphics[width=3.0in]{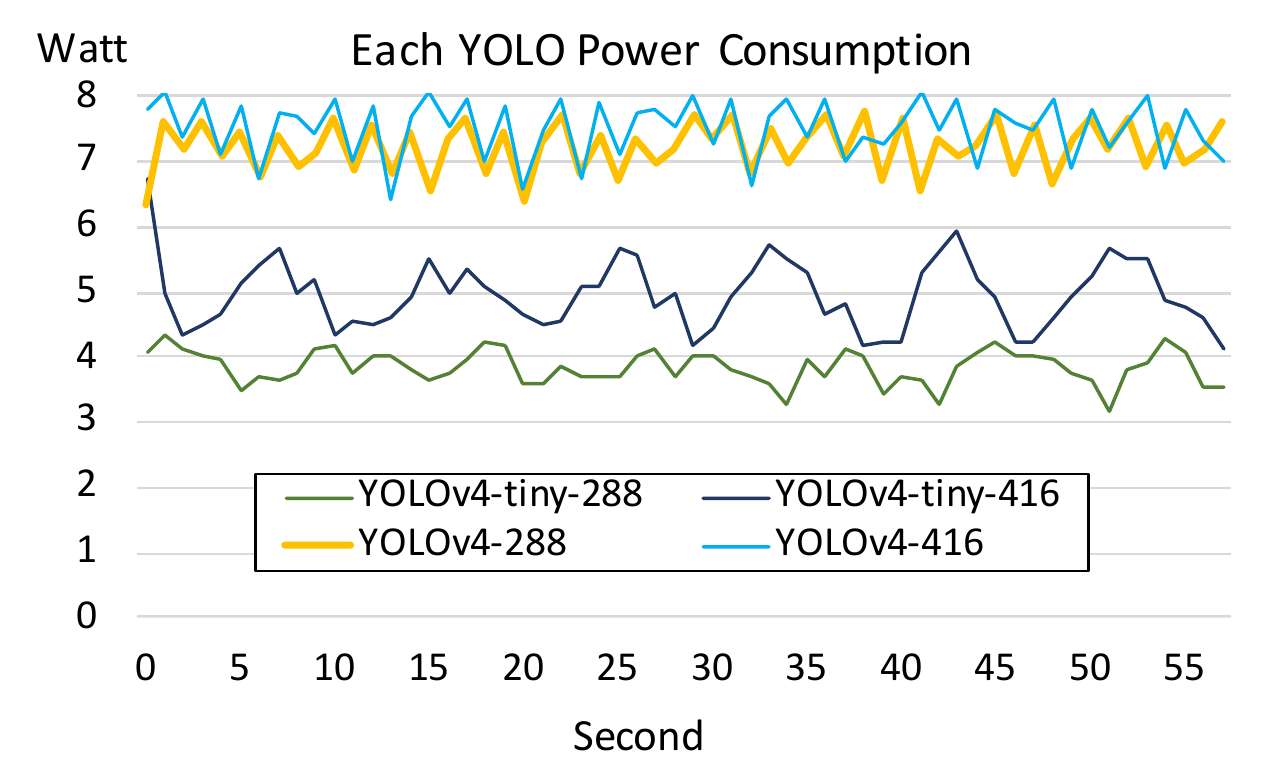}
\caption{Power Consumption on Jetson Nano for individual YOLO using MOT17-05}
\label{fig:power_each}
\end{figure}

\begin{figure}[!t] 
\centering
\includegraphics[width=2.8in]{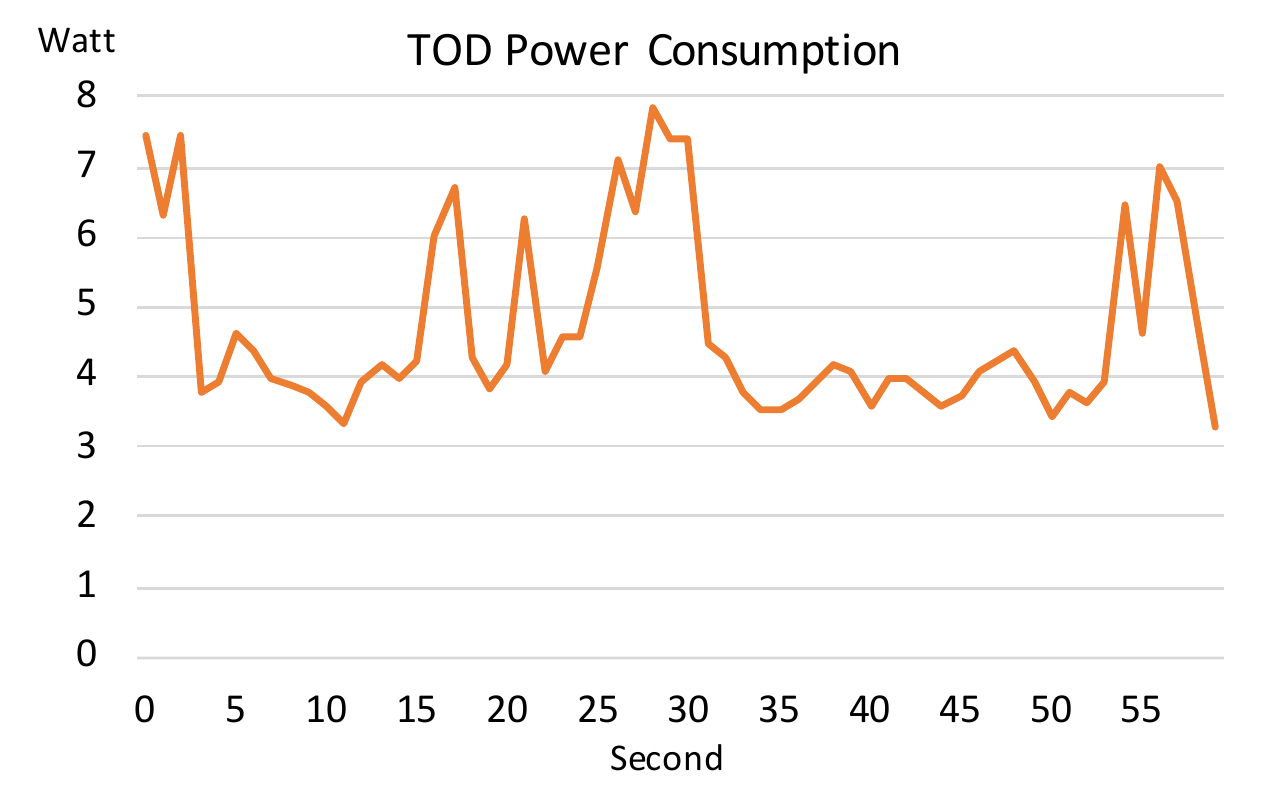}
\caption{Power Consumption on Jetson Nano for TOD using MOT17-05}
\label{fig:power_tod}
\end{figure}

Fig.~\ref{fig:power_each} shows the power consumption of each individual YOLO - 3.8, 4.8, 7.2 and 7.5 $W$ on average for YOLOv4-tiny-288, YOLOv4-tiny-416, YOLOv4-288 and YOLOv4-416 respectively. Fig.~\ref{fig:power_tod} describes the power consumption by TOD. It requires 4.7 $W$ on average to run TOD. Between 15 and 30 seconds, TOD requires relatively higher power (5.7 $W$ on average) due to the execution partially from full YOLO models as shown in Fig.~\ref{fig:dnn_usage}. The power consumption for TOD is lower than the two full version YOLOs and YOLOv4-tiny-416. \emph{Our TOD requires $62.7 \%$ of the GPU board power without losing accuracy, compared to YOLOv4-416.  } 


\section{Discussion} \label{sec:discussion}

\textit{Comparison to Related Work}:
Our approach is more efficient for object detection on the edge when compared to \cite{jiang-chameleon}. Rather than seeking the optimal FPS a DNN can handle~\cite{jiang-chameleon}, TOD automatically chooses a slower network if a higher FPS is not required. This is achieved by using the optimal set chosen by the hyperparameter search technique used in machine learning applications to fit a model on a given dataset \cite{kohavi-cv}. 
This reduces the computational burden by avoiding the choice of an optimal FPS made by the periodic assessments using a heavyweight DNN, compared to \cite{jiang-chameleon}. Moreover, it avoids a profiling method, which has a significant computational overhead. TOD has minimal computational cost since it exploits the specific object detection algorithm characteristics found in \cite{huang-speed} and \cite{kohavi-cv}. Compared to the image classification approach in~\cite{marco-optimizing}, our work focuses on object detection. 

\textit{Hyperparameter Variation Depending on Use Cases}:
Our hybrid DNN TOD is suitable for pedestrian detection on a Jetson Nano, since $H_{opt}$ is chosen under pedestrian datasets given inference latency on a Jetson Nano from each DNN. If TOD is used for another dataset or on another computing platform, $H_{opt}$ may be different since the hyperparameter search returns the best hyperparameter set given the setup. For example, if we utilise a device GPU such as a RTX2080i, the inference latency will be shorter than the Jetson Nano. With less dropped frames from full version YOLOs, the hyperparameter search might return a $H_{opt}$ removing all of the YOLO-tiny version DNNs. YOLO-tiny version DNNs are more useful for resource limited computing platforms such as edge devices. Based on our experiments, full YOLO version DNNs are used more frequently than tiny YOLO version DNNs in TOD since the dropped frames from a Jetson Nano do not impact the accuracy significantly for MOT17Det datasets. If a user is interested in detecting cars on a highway, the hyperparameter search will return the most suitable model for detection. For example, a greater deployment frequency of DNN usage can be assigned to YOLO-tiny DNNs since cars move faster than pedestrians. 

TOD shows minor accuracy loss for MOT17-05, MOT17-11, and MOT17-13 datasets, since hyperparameter search considers the best average prediction for a static camera, a camera moving at walking speed, and a camera moving at the speed of a car. However, TOD achieves equivalent accuracy to the best accuracy over all datasets in MOT17Det.

\section{Conclusions} 
\label{sec:conclusion}

TOD keeps the equivalent accuracy to the best accuracy out of individual DNNs and 
improves the average precision by $34.7, 7.0, 3.9, 2.0 \%$, compared to YOLOv4-tiny-288, YOLOv4-tiny-416, YOLOv4-288 and YOLOv4-416 respectively on average over all MOT17Det datasets. TOD utilises less GPU resource with lower power than YOLOv4-416 without loss in accuracy with the MOT17-05 test dataset. It is envisaged that our TOD can allow more full processing of real-time object detection on edge devices and thus contribute to minimising the amount of data that needs to be sent to the cloud. The longer term focus is to extend TOD to scheduling distributed streaming applications in fog and edge computing environments to maximise either accuracy or energy efficiency.  

\section*{Acknowledgment}
This project has received funding by the European Commission Horizon 2020 research and innovation programme under the grant agreement No. 732631 (OPRECOMP), by the Engineering and Physical Sciences Research Council under the grant agreement No. EP/T022345/1 and by CHIST-ERA under the grant agreement No. ANR-19-CHR3-0002 (DiPET).
We also would like to thank Jesus Martinez del Rincon, Yang Hua, Cheol-Ho Hong and Umar Minhas for their input.


\end{document}